\documentclass[12pt,preprint]{aastex}

\shorttitle{Gravitational Waves of Jet Precession}
\shortauthors{Sun et al.}
\usepackage{textcomp}
\usepackage{amsmath}
\usepackage{graphicx}

\begin{document}

\title{Gravitational Waves of Jet Precession in Gamma-ray Bursts}

 \author{Mou-Yuan Sun\altaffilmark{1},
 Tong Liu\altaffilmark{1},
 Wei-Min Gu\altaffilmark{1,2},
 and Ju-Fu Lu\altaffilmark{1}}

\altaffiltext{1}{Department of Physics and Institute of Theoretical Physics
and Astrophysics, Xiamen University, Xiamen, Fujian 361005, China}

\altaffiltext{2}{Harvard-Smithsonian Center for Astrophysics,
60 Garden Street, Cambridge, MA 02138, USA}

\email{ tongliu@xmu.edu.cn }

\begin{abstract}
The physical nature of gamma-ray bursts (GRBs) are believed to involve
an ultra-relativistic jet. The observed complex structure of light
curves motivate the idea of jet precession.
In this work, we study the gravitational waves of jet precession
based on neutrino-dominated accretion disks around black holes,
which may account for the central engine of GRBs.
In our model, the jet and the inner part of the disk may precess
along with the black hole, which is driven by the outer part of the disk.
Gravitational waves are therefore expected to be significant
from this black hole-inner disk precession system.
By comparing our numerical results with the sensitivity
of some detectors,
we find that it is possible for DECIGO and BBO to detect
such gravitational waves, particularly for GRBs in the Local Group.
\end{abstract}


\section{Introduction}

Gamma-ray bursts (GRBs), which were first detected in 1967, are intense flashes
of gamma rays followed by fainter multiple-wavelength (X-ray, optical, radio)
emission named ``afterglow". Physically, GRBs are believed to be emitted
from ultra-relativistic jets. The afterglow emission is also expected as the
jet interacts with the ambient medium  \citep{Piran04}. One popular
central engine model that powers GRBs consists of a black hole
and a neutrino-dominated accretion flow (NDAF) with mass accretion rates
around $0.01 \sim 10\,\rm{M_{\odot}\ s^{-1}}$ \citep[see,
e.g.,][]{PW99,NPK01,KNP05,LRP05,Gu06,CB07,Liu07,Liu08,Liu12,Lei09}.
Detailed studies have revealed that the quasi-periodic structure
exists in a wide variety of observed light profiles of GRBs
\citep{Romero99,Liang04}. One possible explanation
(particularly for those with a fast rise and exponential decay, i.e., ``FRED")
involves a precessing jet
\citep[e.g.,][]{Bman96,PSLL99,Rey06,Lei07,Liu10}.

\cite{Bman96} investigated the precession of a relativistic blob-emitting
neutron star jet in a binary pulsar (PSR-binary).
They considered Newtonian tidal
torque and gravitomagnetic interaction between PSR-binary to account
for precession and nutation frequencies. Later, \cite{PSLL99} studied a
black hole-neutron star model for GRBs by the black hole forcing the
accretion disk and the jet to precess due to the Newtonian tidal torque.
Moreover, their study showed that the theoretical light curves can fit
the observational data well if one considers the modulation of jet precession on
intrinsic light curves.
The gravitomagnetic interaction between the black hole and the accretion disk is,
however, much stronger than the tidal force \citep{BH}. \cite{Rey06}
suggested the gravitomagnetic interaction as a new basis for
jet precession in GRBs.
The size of the accretion disk needs to be small enough to produce a precession
period $\sim 0.5\,\rm{s}$,
since in their model the whole disk precesses around the black hole.
Furthermore, \cite{Lei07} suggested that the modulated evolution of
the half-opening angle may play the role,
responsible for light curves of GRBs.

Another jet precession model in GRBs was proposed by \cite{Liu10}, which
followed the previous scheme in X-ray binaries \citep[e.g.,][]{Sa80}
and active galactic nuclei \citep[e.g.,][]{Lu90,Lu05}.
\cite{Liu10} argued that the outer neutrino-dominated accretion disk,
whose angular momentum is sufficiently larger than that of the black hole,
can maintain its orientation and force the black hole to precess.
On the contrary, the inner disk whose angular momentum
is significantly smaller than
that of the black hole, should be aligned with
the black hole spin axis \citep{BP75}. Hence, the whole dynamical system
becomes a precessing black hole-inner disk system.
The ultra-relativistic jet, which is launched by the inner disk
aligned with the black hole spin axis, will also precess and produce
the observational complex structure in the light curves.

Apart from the electromagnetic emission in GRBs,
gravitational waves are also expected due to the black hole-inner disk
precession.
Different from gamma-ray and afterglow emission, which is believed
to be produced far from the black hole, the gravitational radiation
should be emitted near the central engine.
On the other hand, many efforts have been made to
detect gravitational wave signals with some current detectors such as LIGO
and some still undergoing detectors like DECIGO, BBO, and LISA.
The main purpose of this work is to study the detectability
of the above mentioned gravitational waves.

This paper is organized as follows. In Section 2 we describe the
disk-driven jet precession model. In Section 3 we show the formulae for calculating
gravitational waves emitted by the black hole-inner disk
precession system. In Section 4 we present our numerical results and
discuss the possibility of detecting such gravitational waves by some
future detectors. Conclusions and discussion are made in Section 5.

\section{Disk-Driven Jet Precession Model}
\label{s2}
In this section, we review the main features of the disk-driven jet
precession model \citep[for details, see][]{Liu10}.
The model we considered is a Kerr black hole surrounded by
a tilted accretion disk whose initial orbital axis is misaligned
with the black hole spinning axis.
The angular momentum per each ring at radius $r$ with
width $dr$ is $dJ=2\pi r^2\Sigma v_{\varphi}dr$,
where $\Sigma$ is the surface density and
$v_{\varphi}$ is the azimuthal velocity.
A typical angular momentum of the
disk is \citep[e.g.,][]{Sa80}
\begin{equation}
J = \frac{dJ}{d(\ln r)} = 2\pi r^3\Sigma v_{\varphi}\ .
\end{equation}
There exists a critical radius $r_{\rm{p}}$ where
the typical angular momentum $J|_{r=r_{\rm p}}$ is equal to
the black hole angular momentum $J_\ast$, i.e.,
\begin{equation}
\label{Jrp}
J|_{r=r_{\rm p}} = J_{\ast} = \frac{GM^2 a_{\ast}}{c} \ ,
\end{equation}
where $M$ is the black hole mass and $a_\ast$ is a dimensionless
spin parameter ($0<a_\ast<1$).
The outer disk ($r>r_{\rm{p}}$)
will maintain its orientation and therefore force the black hole and
the inner disk ($r<r_{\rm{p}}$) to be a whole precessing system.
The precession rate is expressed as \citep[e.g.,][]{Sa80,Lu90}
\begin{equation}
\label{ome}
\Omega=\frac{2GJ_{\ast}}{c^2r_{\rm p}^3}\ .
\end{equation}
The mass conservation equation takes the form,
\begin{equation}
\dot{M}=-2\pi r\Sigma v_{r}\ ,
\end{equation}
where $\dot{M}$ is the mass accretion rate and $v_{r}$ is
the radial velocity of the flow.
The precession period can be derived by combining Eqs.~(1)-(4)
to eliminate $\Sigma$, $r_{\rm p}$, and $J_{\ast}$,
\begin{equation}
\label{PP}
P = \frac{2\pi}{\Omega}
= \pi M (\frac{a_\ast}{G})^{\frac{1}{2}}(-\frac{cv_{r}|_{r=r_{\rm p}}}
{\dot{M}v_{\varphi}|_{r=r_{\rm p}}})^{\frac{3}{2}}\ .
\end{equation}

Following \cite{RH95}, we introduce the general relativity correction as
\begin{equation}
A=1-\frac{2GM}{c^2r}+(\frac{GM a_{\ast}}{c^2r})^2\ ,
\end{equation}
\begin{equation}
B=1-\frac{3GM}{c^2r}+2a_{\ast}(\frac{GM}{c^2r})^{3/2}\ ,
\end{equation}
\begin{equation}
C=1-4a_{\ast}(\frac{GM}{c^2r})^{\frac{3}{2}}+3(\frac{GMa_{\ast}}{c^2r})^2\ ,
\end{equation}
\begin{equation}
D=\int^r_{r_{\rm{ms}}}\frac{x^2c^4/(2G^2)-3xMc^2/G+4(xa_{\ast}^2M^3c^2/G)^{1/2}-3a_{\ast}^2M^2/2}{(xr)^{1/2}[x^2c^4/G^{2}-3xMc^2/G+2(xa_{\ast}^2M^3c^2/G)^{1/2}]}dx\ ,
\end{equation}
where $r_{\rm{ms}}$ is the radius of inner marginally stable orbit.
The hydrostatic balance in $z$-direction takes the form:
\begin{equation}
\frac{1}{\rho}\frac{\partial p}{\partial z} = - \frac{GMz}{r^3}\frac{C}{B}\ ,
\end{equation}
where $\rho$ is the mass density, and $p$ is the total pressure.
We make an improvement on describing the vertical structure
by a polytropic relation, i.e., $p=K\rho^{1+1/N}$, instead of
the one-zone approximation in \cite{Liu10},
where $1+1/N$ is the polytropic index.
The above hydrostatic equation then gives \citep[e.g.,][]{H77}
\begin{equation}
\rho(r,z)=\rho_0(r)(1-\frac{z^2}{H^2})^N\ ,
\end{equation}
\begin{equation}
p(r,z)=p_0(r)(1-\frac{z^2}{H^2})^{N+1}\ ,
\end{equation}
where $\rho_0$ and $p_0$ are the density and the pressure
on the equatorial plane, respectively.
The half-thickness of the disk, $H$, is expressed as
\begin{equation}
\Omega_{\rm{K}}^2H^2=2(N+1)\frac{p_0}{\rho_0}\frac{B}{C}\ ,
\end{equation}
where $\Omega_{\rm K}$ is the Keplerian angular velocity.
The above equations combined with
the energy and momentum conservation equations and the equation of state
\citep[see, e.g.,][]{Liu07} enable us to solve the structure of
the disk and consequently to obtain the precession period.
Then we can go further to investigate the property of
gravitational waves from such a black hole-inner disk precession system.

\section{Gravitational waves from the black hole-inner disk precession system}
\label{s3}
Here we consider the black hole-inner disk precessing system
as an axisymmetric rigid system(with moments of inertia $I_1=I_2$).
Such a system would precess with a period given in Equation (\ref{PP}).
In the body frame, the inertia tensor is diagonal with
eigenvalues $I_1, I_2, I_3$, where
\begin{equation}
I_3=\int_{r<r_{\rm{p}}}(x^2+y^2)\rho(x,y,z)dxdydz\ ,
\end{equation}
\begin{equation}
I_1=I_2=\int_{r<r_{\rm{p}}}(z^2+y^2)\rho(x,y,z)dxdydz\ .
\end{equation}
The precessing motion of such a rigid system is a classical
Newtonian problem and it is easy to obtain the inertia tensor
in the observing frame \citep[e.g.,][]{LL76,MM08}:
\begin{equation}
I_{xx}=\frac{1}{2}(I_1-I_3)\sin^2\theta\cos(2\Omega t)+C_1\ ,
\end{equation}
\begin{equation}
I_{xy}=\frac{1}{2}\sin^2\theta\sin(2\Omega t)\ ,
\end{equation}
\begin{equation}
I_{yy}=-\frac{1}{2}(I_1-I_3)\sin^2\theta\cos(2\Omega t)+C_2\ ,
\end{equation}
\begin{equation}
I_{xz}=-(I_1-I_3)\sin\theta\cos\theta\sin(2\Omega t)\ ,
\end{equation}
\begin{equation}
I_{yz}=-I_{xz}\ ,
\end{equation}
\begin{equation}
I_{zz}=I_1\sin^2\theta+I_3\cos^2\theta\ ,
\end{equation}
where $\theta$ is the misaligned angle between the black hole spin axis
and the orientation of the outer disk, and $\Omega$ is the precession rate
given by Equation (\ref{ome}). Note that $C_1$ and $C_2$
represent some constants which are unimportant in our calculation since
gravitational waves are only relevant to the time-dependent components
of moments of inertia.

Gravitational waves will be produced since the inertia tensor
in the observing frame is time dependent.
With the assumption that the angle between the $z$-axis of gravitational wave
detector and the signal direction of arrival is $\iota$ and the distance of
the GRB is $d$, the amplitude of gravitational waves is given by
\citep[e.g.,][]{zs79,MM08}

\begin{equation}
h_{\rm{pre}}(t)=h_+(t)+h_{\times}(t)\ ,
\end{equation}
where
\begin{equation}
\begin{split} h_+(t)=h_0\sin 2\theta \cos(\Omega
t)\sin \iota \cos\iota\\
+2h_0\sin^2\theta \cos(2\Omega t)(1+\cos^2\iota)\ ,
\end{split}
\end{equation}
\begin{equation}
\begin{split}
h_{\times}(t)=h_0\sin 2\theta \sin(\Omega t)\sin \iota\\
+4h_0\sin^2\theta \sin(2\Omega t)\cos \iota\ ,
\end{split}
\end{equation}
with
\begin{equation}
h_0=-\frac{G}{c^4}\frac{(I_3-I_1)\Omega^2}{d}\ .
\end{equation}
Thus the black hole-inner disk system emits gravitational waves
at two frequencies, i.e.,
$f_{\rm{gw}}=\Omega/2\pi$ and $\Omega/\pi$.
Meanwhile, since the hyper-accretion in GRBs exists only for
seconds and would finally stop, the gravitational wave is this work
should be a gravitational wave burst and be significant
only when the central engine is active. Thus, the duration of
the gravitational wave burst is roughly equal to the activity time
of the central engine of GRBs.
The gravitational wave signal waveform is therefore expected
to be \citep[e.g.,][]{MM08}
\begin{equation}
h(t)=h_{\rm{pre}}(t) e^{-\frac{t^2}{2\delta_{\rm{pla}}^2}}\ ,
\end{equation}
where $\delta_{\rm{pla}}$ is the plateau time of GRBs,
which is roughly the duration of activity time of the central engine.
In order to assess the detectability, we calculate the
root-sum-square (rss) amplitude as follows \citep[e.g.,][]{Ace08,MM08}:
\begin{equation}
h_{\rm{rss}}(f)=\sqrt{\int_{-\infty}^{\infty}(h_+^2(t)+h_{\times}^2(t))dt}\ ,
\end{equation}
where $f=\Omega/2 \pi=1/P$.

Gravitational waves also carry energy and momentum. The quadrupole power of gravitational wave is
\begin{equation}
P_{\rm quad}=\frac{G}{5c^5}\left\langle \dddot{M_{ij}} \dddot{M_{ij}}-\frac{1}{3}(\dddot{M_{kk}})^2\right\rangle \ .
\end{equation}
In our case, the momentum $M_{ij}$ satisfies the relation $\dddot{M_{ij}}=\dddot{I_{ij}}$ \citep[e.g.,][]{MM08}.
Therefore, the quadrupole power radiated is
\begin{equation}
P_{\rm quad}=\frac{2G}{5c^5}(I_{1}-I_{3})^2\Omega^6\sin^2\theta(1+15\sin^2\theta)\ .
\end{equation}

\section{Numerical Results}

The structure of our jet precession model is determined by
$\dot{M}$, $a_\ast$, $M$, and the viscosity parameter $\alpha$.
Following \cite{Liu10}, we adopt $\alpha = 0.01$ in our calculation.
In addition, the polytropic index $1+1/N$ is set to be $5/3$.
We then numerically solve the equations described in
Section~\ref{s2} with given $\dot{M}$, $M$, and $a_\ast$ to
obtain the structure of the system. In order to calculate the strength of
gravitational waves, we assume that the misaligned
angle between the black hole spinning axis and the accretion disk orientation is $\theta=\rm{20^{\circ}}$.
We also fix the duration gravitational wave bursts \citep[roughly the
activity time of the central engine of GRBs, which is, see e.g.,][an order of $20\ \rm s$]{PSLL99}
$\delta_{\rm{pla}}=20\ \rm s$ for illustration purpose.
In addition, for given distance $d$ of the GRB source, we can derive the
rss amplitude and quadrupole power of gravitational waves by equations in Section~\ref{s3}.

\subsection{Disk-driven Jet Precession}

\begin{figure}
\plotone{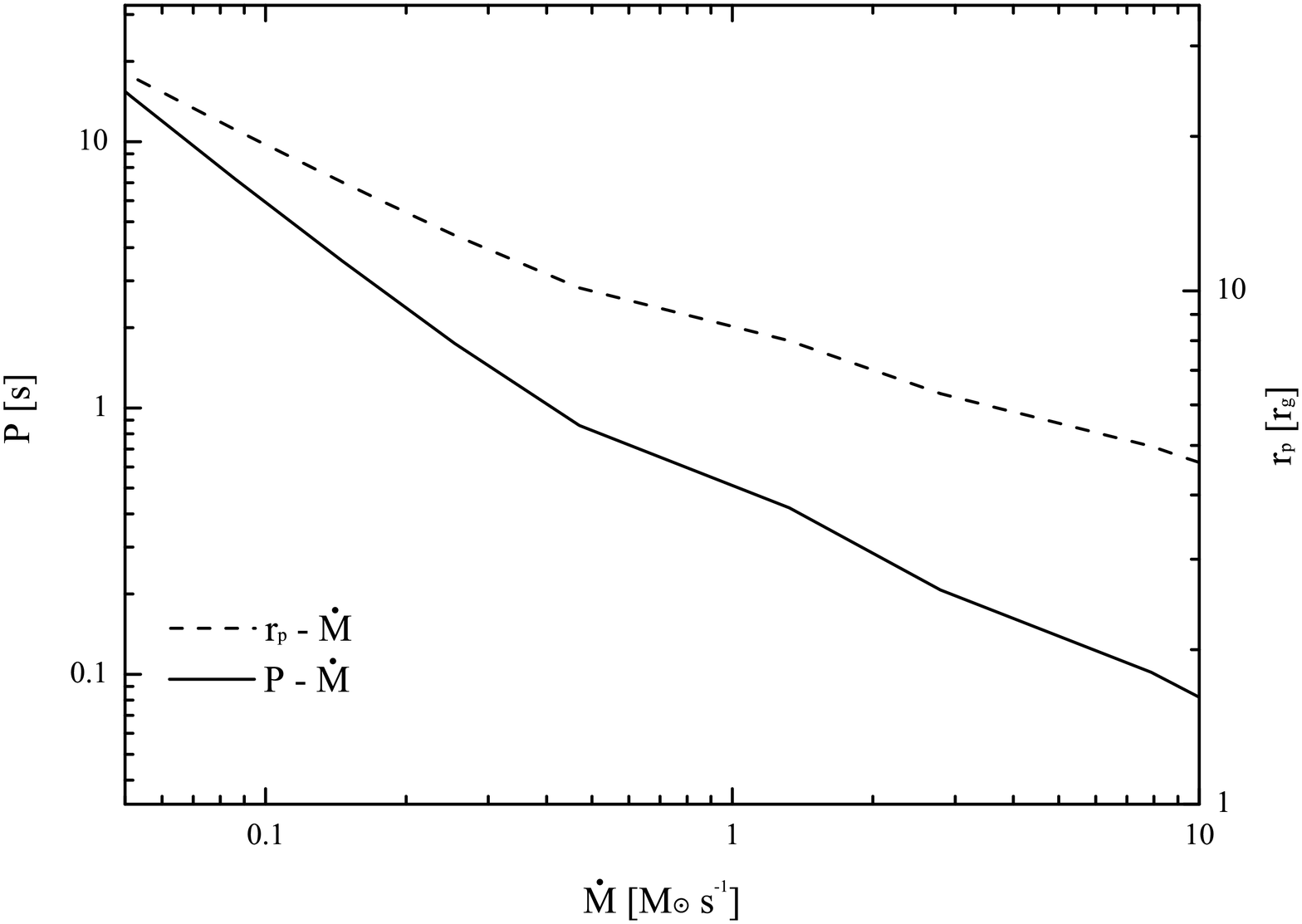}
\caption{Variation of the precession period $P$ (solid line) and critical radius
$r_{\rm p}$ (dashed line) with the accretion rate $\dot{M}$,
for which $M=6 \ \rm{M_{\odot}}$ and $a_{\ast}=0.9$.
}
\label{f1}
\end{figure}

With fixed $M=6\,\rm{M_{\odot}}$ and $\alpha=0.01$,
we illustrate the possible jet precession period $P$ as a
function of the accretion rate (solid line) in Figure~\ref{f1}.
In addition, we plot
the critical radius $r_{\rm p}$ as a function of the accretion rate
(dashed line).
It is seen that NDAFs with $\dot{M}=0.05 \sim 10\,\rm{M_{\odot}\ s^{-1}}$
can drive jet precession with period $P=10 \sim 0.1\,\rm{s}$.
Thus the disk-driven jet precession may explain the temporal
structures in light curves of GRBs.
Note that the corresponding critical radius $r_{\rm p}$ is
close to the horizon of the black hole and the inner precession disk is
thus very small.

\subsection{Gravitational waves and the black hole spin}

It is known that the black hole spin can affect both the structure and
the precession rate.
In this section we study the dependence of the gravitational wave rss
amplitude on the black hole spin for a fixed $M$ and $d$.
We vary the spinning parameter from $a_{\ast}=0.1$
to $a_{\ast}=0.95$ and calculate the corresponding gravitational wave
rss amplitude $h_{\rm{rss}}$.

\begin{figure}
\plotone{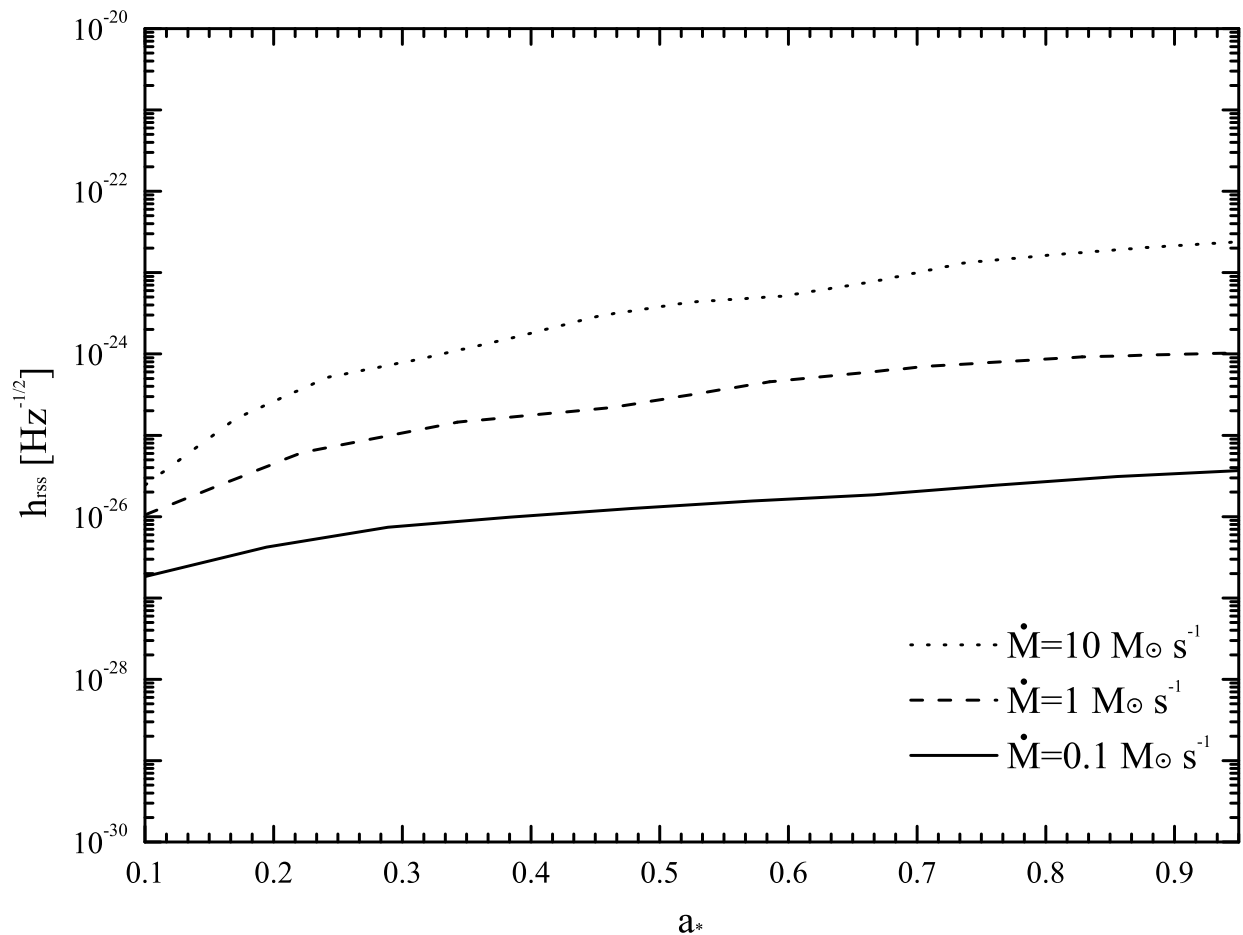}
\caption{Variation of the gravitational wave rss amplitude with
the spin parameter $a_\ast$, for which the black hole mass
$M=6 \ \rm{M_{\odot}}$ and the distance $d=1 \ \rm{Mpc}$.
The solid, dashed, and dotted lines correspond to
the mass accretion rate $\dot{M}=0.1$, $1$, and $10 \ \rm{M_{\odot}\ s^{-1}}$,
respectively.}
\label{f2}
\end{figure}

Figure~\ref{f2} plots the amplitude $h_{\rm rss}$ as a function of
$a_{\ast}$, for which $M=6\,\rm{M_{\odot}}$ and $d=1\,\rm{Mpc}$.
The solid, dashed, and dotted lines correspond to
$\dot{M} = 0.1,\, 1,$ and $10\,\rm{M_{\odot}}\ s^{-1}$, respectively.
It is seen that $h_{\rm rss}$ rapidly increases with $a_{\ast}$
in the low-spin region, whereas $h_{\rm rss}$ becomes flat
in the high-spin region.
This result is easy to understand since the black hole with higher spin
will precess along with a larger inner disk, and consequently
the moments of inertia of the precession system will be larger, which
will result in larger $h_{\rm rss}$.

The formation of an ultra-relativistic jet may require a rapid spinning
black hole \citep[e.g.,][]{NM12}. The hyper accretion process would also spinning up the central
black hole. Therefore, we fix $a_{\ast}=0.95$ for the later calculation
to focus on the variation of $h_{\rm rss}$ and frequency with varying $\dot M$, $M$, and $d$.

\subsection{Gravitational waves and the precession rate}

For a given mass of the black hole, the disk structure and the precession
period are determined only by $\dot M$ ($\alpha$ and $a_{\ast}$
have been fixed), so there exists a certain $P$ (or $f=1/P$) corresponding to
each $\dot M$. In this case, if the distance is also given, then
there also exists a certain $h_{\rm rss}$ corresponding to each
$\dot M$. In this section, we investigate the gravitational waves
for each $\dot{M}$ (and $P$, $f$).

\begin{figure}
\plotone{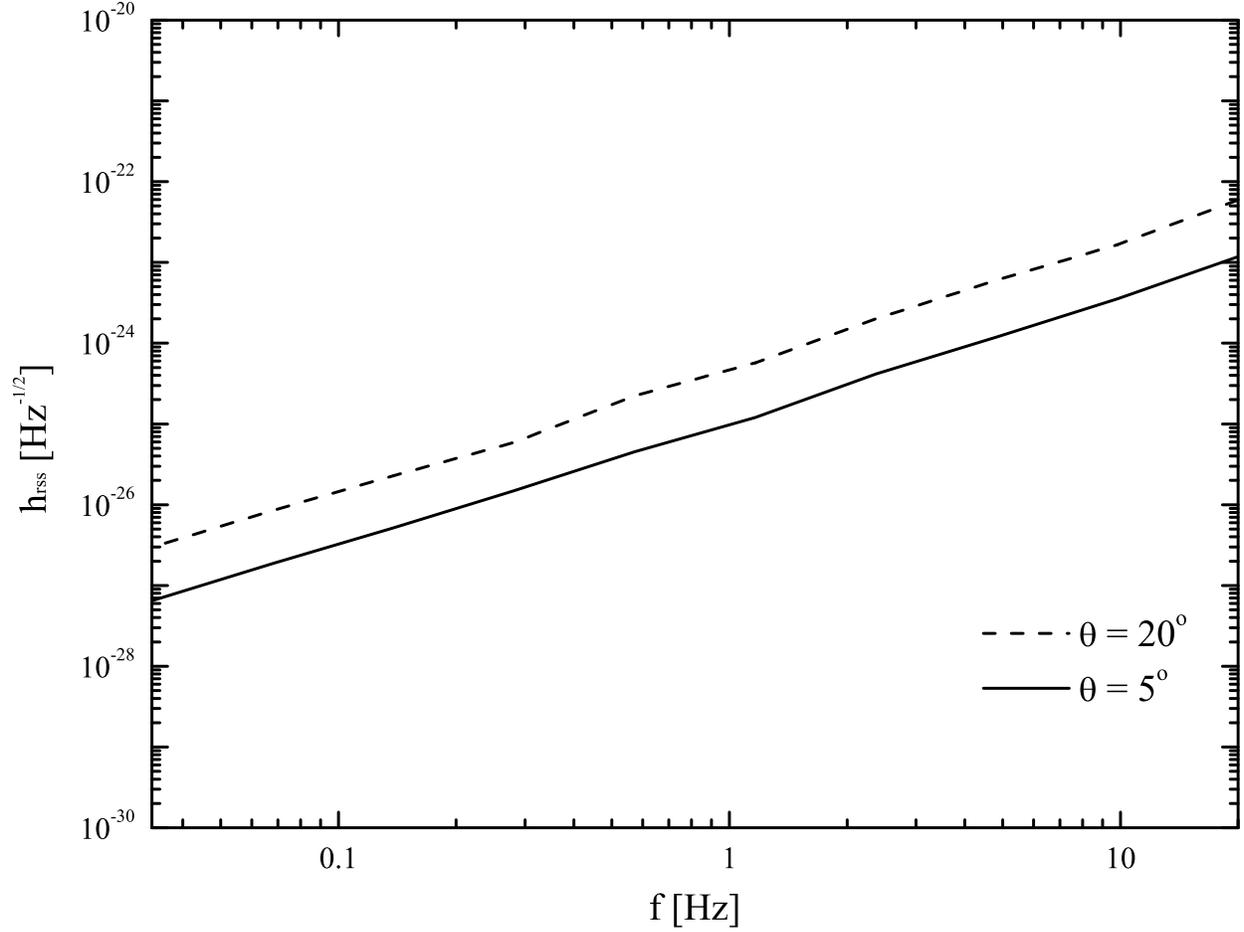}
\caption{The gravitational wave rss amplitude as a function of the frequency
for which $M=6 \ \rm{M_{\odot}}$ and $d=1\,\rm{Mpc}$.
The solid and dashed lines correspond to $\theta=\rm{5^{\circ}}$
and $\theta=\rm{20^{\circ}}$, respectively.}
\label{f3}
\end{figure}

As mentioned at the beginning of this section, we set
$\theta=\rm{20^{\circ}}$ to assess the rss amplitude of gravitational waves since as
shown by \cite{Rey06},  $\theta$ is an order of $\rm{20^{\circ}}$ in some GRBs.
However, $\theta$ may be different for different GRBs.
In Figure~\ref{f3} we present the rss amplitude of
gravitational waves as a function of $f$ for
$\theta=\rm{5^{\circ}}$ (solid line) and $\theta=\rm{20^{\circ}}$
(dashed line). It is seen that the rss amplitude decreases by a factor
of 2 to 3 as $\theta$ varies from $\rm{20^{\circ}}$ to $\rm{5^{\circ}}$.
We will keep $\theta=\rm{20^{\circ}}$ for calculation
in the remainder of this paper.

Figure~\ref{f4} shows the relationship among $h_{\rm rss}$, $f$, and
$\dot M$, for which $M=6\,\rm{M_{\odot}}$. The range of $\dot M$ is roughly
$0.01 \sim 10 \ \rm{M_{\odot}\ s^{-1}}$, which is known as the possible
$\dot M$ for GRBs.
The dash-dotted, solid, and dashed lines correspond to the results for
$d=10\,\rm{kpc}$, $1\,\rm{Mpc}$, and $100\,\rm{Mpc}$, respectively.
The dotted lines represent the detectability of DECIGO, BBO, and
the ultimate DECIGO \citep[e.g.,][]{YS11}.
The figure indicates that, for GRBs in the Local Group
($d \la 1\,\rm{Mpc}$), such gravitational waves are under
DECIGO and BBO's detectability.
The event rate is, however, quite low in the Local Group.
For larger distance $d \la 100\,\rm{Mpc}$,
the event rate may increase to $\sim 1\,\rm{yr^{-1}}$,
but the rss amplitude $h_{\rm rss}$ is too small to be detected by practical DECIGO or BBO.
Nevertheless, as shown in this figure, such gravitational waves are under
detectability of the ultimate DECIGO, which roughly represents
the quantum level noise of gravitational wave detectors.

It is also seen from Figure~\ref{f4} that
the amplitude $h_{\rm{rss}}$ and the frequency $f$ both increase with
$\dot M$. This is because higher accretion rates generally
correspond to higher mass density and therefore larger moments of inertia
and larger gravitational wave rss amplitude. On the other hand,
the critical radius $r_{\rm{p}}$,
will become smaller for higher accretion rates, which results in higher
precession rates.

We also study the strength of such gravitational waves for various
central black hole masses. Figure~\ref{f5} presents the variation
of $h_{\rm rss}$ with $f$ for a fixed distance $d = 1\,\rm{Mpc}$.
The dashed, solid, and dash-dotted lines correspond to the
black hole mass $M=3,\, 6$, and $10\,\rm{M_{\odot}}$, respectively.
It is seen that $h_{\rm{rss}}$ increases with the
black hole mass, which can be understood as follows. The critical radius
$r_{\rm p}$ increases with the black hole angular momentum
$J_{\ast}$ (or $M$ since $a_{\ast}$ is fixed), so the larger $M$
corresponds to the larger moments of inertia of the inner disk, and therefore
stronger gravitational waves. We would like to point out that,
even though larger black hole may produce stronger gravitational wave rss amplitude,
$h_{\rm rss}$ cannot be essentially enhanced since the
black hole mass in GRBs is limited, normally $M \la 10\,\rm{M_{\odot}}$.

\subsection{The power of gravitational wave}
In order to further explore the gravitational wave from our model, we calculate the quadrupole
power of gravitational waves $P_{\rm quad}$. Figure~\ref{f6} plots $P_{\rm quad}$ as a function of $f$, for which
$M=6\,\rm{M_{\odot}}$ and $a_{\ast}=0.95$.
In addition, the luminosity range of GRBs is shown as the shaded region \citep[e.g.,][]{ZB11}.
It is seen that the power of the gravitational wave is significantly less than the isotropic luminosity
of the GRBs. The gravitational potential energy released in the accretion disk
is mainly converted into the neutrino radiation rather than
the gravitational radiation. Therefore, our assumption that the structure of NDAFs
is not affected by the production of gravitational waves should be self-consistent.

\section{Conclusions and Discussion}

In this paper, we have studied the gravitational waves from
GRBs with disk-driven jet precession. Based on the model
in \cite{Liu10}, we have calculated gravitational wave
rss amplitude and obtained the variation of the
amplitude with the frequency for different values of $M$ and $d$.
By comparing our numerical results with the sensitivity
of some detectors, we have found that it is possible for
DECIGO and BBO to detect such gravitational waves, particularly for
GRBs in the Local Group.

The rate of GRBs in the Local Group is apparently low.
Hence, the detection rate of gravitational waves by our model is
quite low \citep[see][for a detailed discussion on the gravitational wave
event rate that associates with GRBs]{L09}.
However, such gravitational waves may still be detected
from systems absence of GRB events.
On one hand, there is strong evidence showing that
X-Ray Flashes (XRFs) and GRBs
are just two types of bursts with the same physical nature \citep[e.g.,][]{Lamb03}.
On the other hand, observational and theoretical arguments both
indicate that the so-called ``failed'' GRBs exist if the jet is dirty
(baryon-rich) and cannot breakthrough the envelope \citep[e.g.,][]{HDL02,TO03}.
Meanwhile, we argue that disk-driven jet precession may be common in black
hole accretion system since the only necessary condition is that
the angular momentum of the initial accretion flow is misaligned
with the black hole spinning axis.
Thus the similar gravitational waves may also be produced both in
``failed'' GRBs and XRFs. Then the detection rate is probably
related to the total rate of ``failed'' GRBs and XRFs and
therefore increase significantly.
For example, if the total rate
approaches the SNe Ib/c event rate (of course this should just be
regarded as an upper limit), the expected detection rate by
our model will increase to
$10^{-2}\sim 10^{-1}\ \rm{yr^{-1}}$
\citep[for the SNe Ib/c rate, see, e.g.,][]{Pod04}.
More importantly, the gravitational wave signals may be the unique way
to explore the nature of off-axis GRBs and ``failed'' GRBs
except for the possible orphan afterglow emission.
For example, the trigger time of the orphan afterglow can be measured
if the above mentioned gravitational waves are detected.
Then one can make a distinguish between off-axis GRBs
and ``failed'' GRBs by studying the time evolution of the orphan
afterglow \citep[see details in][]{HDL02}.

Some studies \citep[e.g.,][]{SINY,HKKT,SM09} also focused on the low frequency
gravitational waves (typically, $f \lesssim 10\ \rm Hz$) emitted from GRBs.
\cite{SINY} studied the gravitational waves from the acceleration stage of
GRB jets based on the internal shock model. The gravitational waves they studied
have a ``memory effect" (that is, gravitational waves would survive at the end of the
jet acceleration stage) and do not depend on the energy form of jets (that is, whether
the jet is powered by BZ process or neutrino annihilation). \cite{HKKT} considered
that the gravitational waves with ``memory effect" from the neutrino-driven GRB jets.
Such gravitational waves are expected to be stronger than that of \cite{SINY}
since the neutrino luminosity is much higher than the energy released
by matter in jets. \cite{SM09}, however, suggested gravitational waves with ``memory effect"
would generated because of anisotropic neutrino emission above the NDAF in GRBs. All these
gravitational waves can be detected by the gravitational wave detectors such as LISA and
DECIGO if the frequency is less than $1\ \rm Hz$ and the source is located at a few Mpc
\citep[note that for][gravitational waves whose frequencies $f\sim 100\ \rm Hz$ can also
be detected by LIGO]{SM09}. There are many common features between our model and these studies.
For example, gravitational waves are both generated near the central engine \citep[especially for our model and that of][]{SM09},
frequencies of gravitational waves are both very low and the detectable distances
are both about several Mpc. One obvious difference is that in our model gravitational waves
whose frequencies $f=1\sim 10\ \rm Hz$ are more likely to be detected. Therefore, a simultaneously
detection of all these type of gravitational waves may give us a new sight into the central engine of
GRBs.

Other types of gravitational waves emitted from the central engine of GRBs have been
studied by \cite{van03} and \cite{Romero10}. The former suggested that
an inner GRB engine consists of a Kerr black hole surrounded by
a uniform magnetized torus, whose accretion is suspended
because the black hole-torus interaction can
transfer angular momentum to the torus or the disk and
prevent the accreted materials from falling into the horizon.
The instability develops in the black hole-torus system
because magnetic fields break up the axisymetry and a large fraction of
energy of this system is released by gravitational waves, which make GRBs
to be the most powerful gravitational radiation sources. Obviously,
the gravitational waves studied in this work are systematically lower
than that of \cite{van03} both on the amplitude and the frequency
\citep[the frequency of gravitational waves suggested by][is
a few hundred $\rm{Hz}$]{van03}.
On the other hand, gravitational waves in \cite{van03} that powered by the spinning black hole
would slow down the spin and change the structure of the surrounding torus significantly. In this
work, however, as Figure~\ref{f6} shown, the power of gravitational waves is much lower than that
of \cite{van03}. Gravitational waves in this work almost have no affect on the structure of NDAFs.
We can study the structure of NDAFs and the production of gravitational waves separately.

\begin{figure}
\plotone{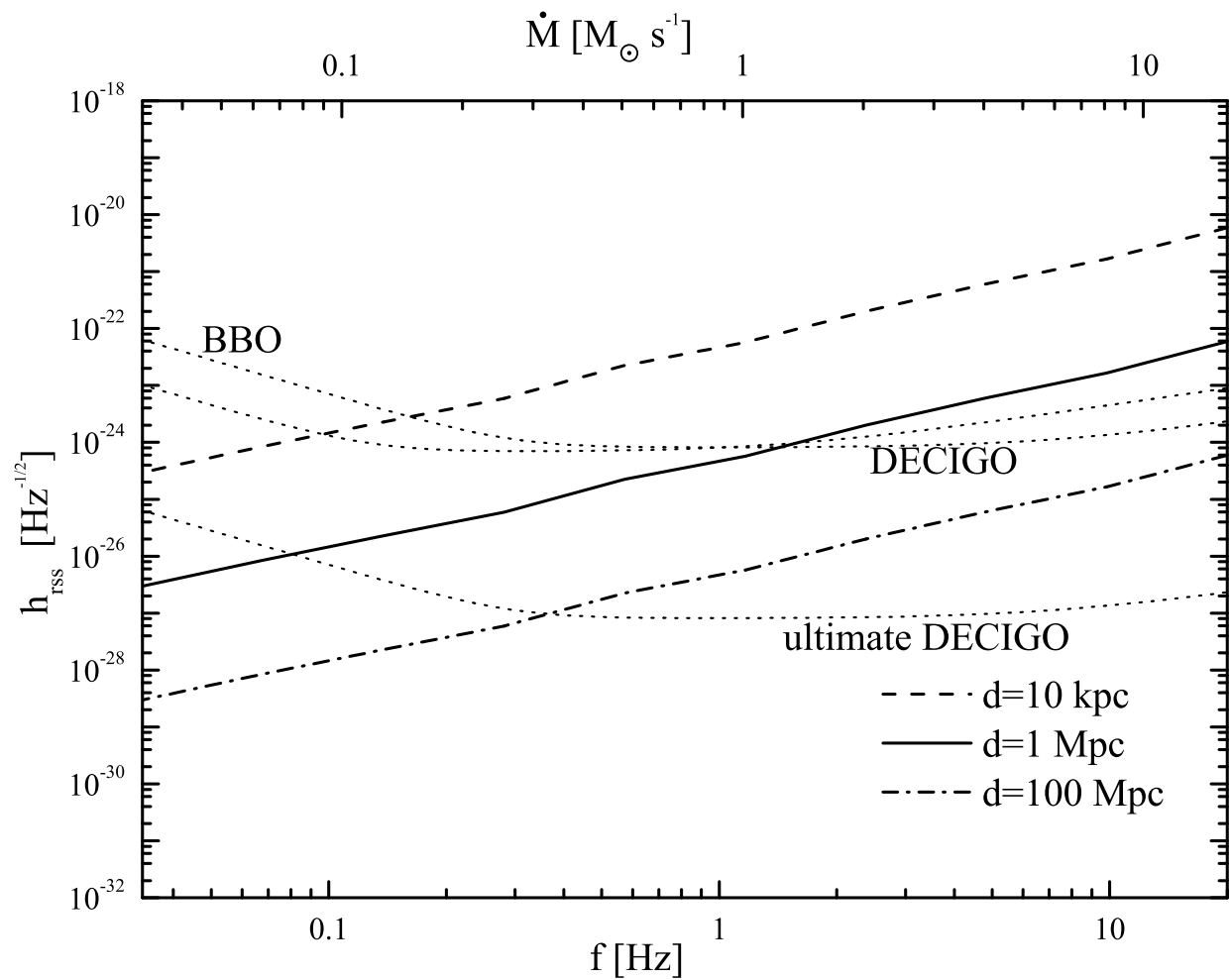}
\caption{The gravitational wave rss amplitude as a function of the frequency
(or the accretion rate),
for which $M=6 \ \rm{M_{\odot}}$. The dashed, solid, and dash-dotted
lines correspond to $d=10\,\rm{kpc}$, $1\,\rm{Mpc}$, and $100\,\rm{Mpc}$,
respectively. The dotted lines represent the detectability of
gravitational wave detectors.}
\label{f4}
\end{figure}

\begin{figure}
\plotone{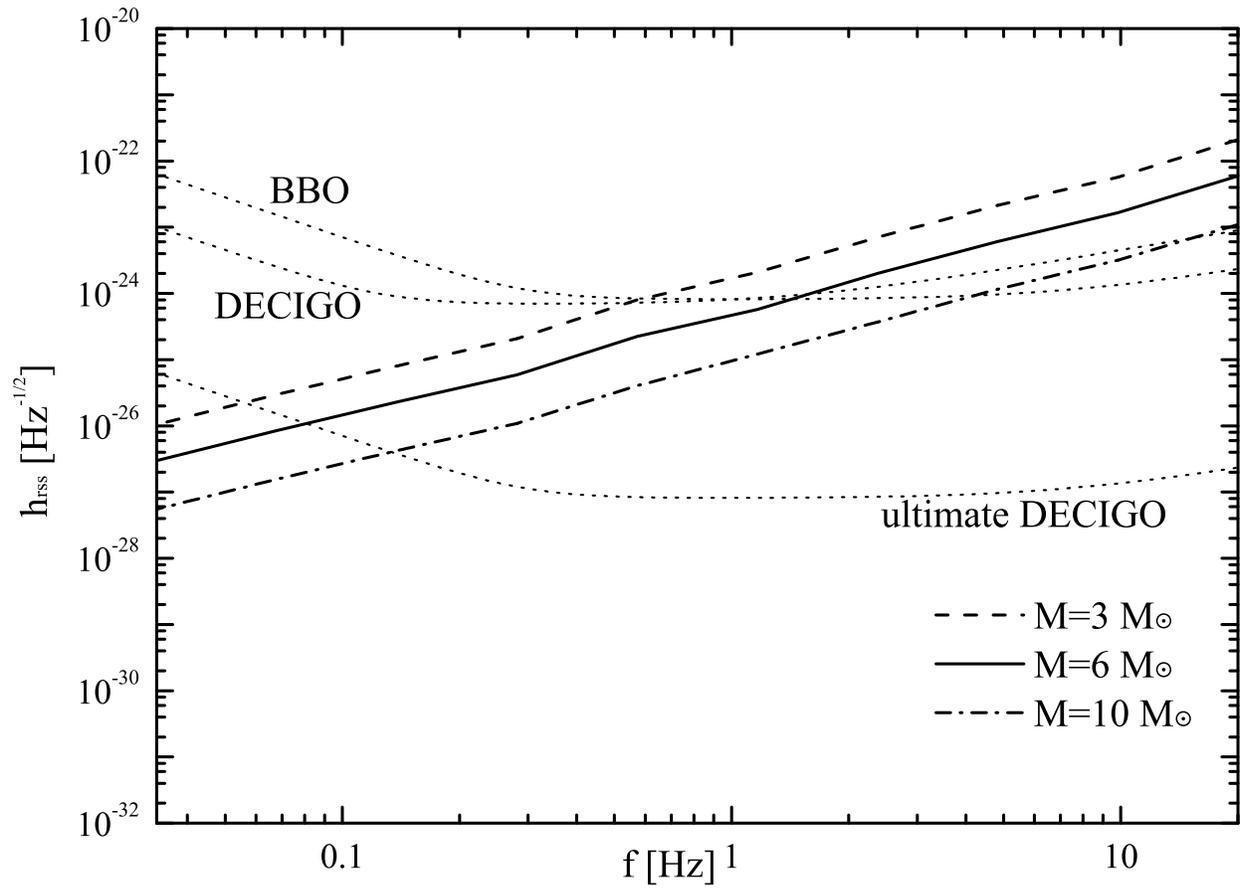}
\caption{Same as Figure~\ref{f2}, except for
$M=3$, $6$, and $10\,\rm{M_{\odot}}$ with a fixed $d=1\,\rm{Mpc}$.}
\label{f5}
\end{figure}

\begin{figure}
\plotone{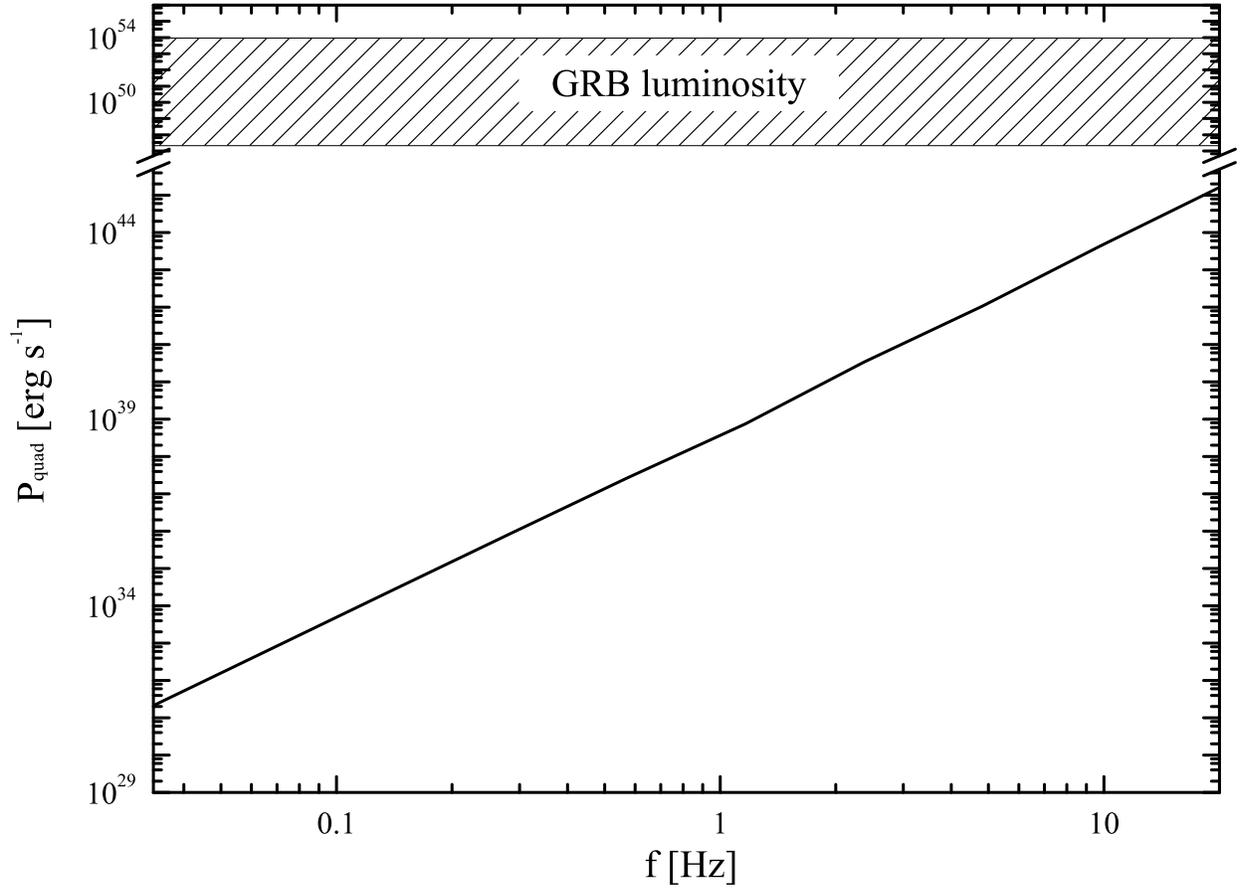}
\caption{The quadruple power of gravitational wave radiated from the
precessing central engine as a function of the frequency $f$, for which
$M=6 \ \rm{M_{\odot}}$.
The shaded region represents the luminosity range of GRBs.}
\label{f6}
\end{figure}

Moreover, \cite{Romero10} considered a new type of
gravitational wave by assuming that the whole
accretion disk precesses as a rigid body. They showed
that gravitational waves from such a precessing system
can be detected by advanced LIGO in the near future (e.g.,
Figure~5 in their paper).
Since the calculation of gravitational wave emission in \cite{Romero10} is
relevant to the formulae developed for a torque-free precession system, the frequency in
their model is significantly different from that in the present work.
Furthermore, the rss amplitude of gravitational waves of \cite{Romero10} should,
in principle, decrease with increasing frequency since the size of the accretion disk
decreases with increasing frequency (e.g., Figure~5 in their paper).
On the contrary, the rss amplitude
in our model increases with frequency as shown in Figures~\ref{f4} and
\ref{f5}.
Such a difference may help distinguish our model from theirs.

\acknowledgments
We thank the anonymous referee for very useful suggestions and comments.
We thank Ye-Fei Yuan, Matias M. Reynoso, and Gustavo E. Romero for beneficial
discussions. This work was supported by the National Basic Research Program
(973 Program) of China under grant 2009CB824800, and the National Natural
Science Foundation of China under grants 10833002, 11073015, and 11103015.

\clearpage

\end{document}